\documentclass[11pt]{article}

\usepackage{amsmath}
\usepackage{amsfonts}
\usepackage{amssymb}
\usepackage{amsthm}
\usepackage{graphicx}
\usepackage{marvosym}
\usepackage{tikz}
\usetikzlibrary{arrows,shapes,positioning}
\usetikzlibrary{calc}
\usepackage{pgfplots}
\usepackage{cleveref}
\usepackage{graphicx}
\usepackage{subcaption}
\usetikzlibrary{trees,snakes}
\usepackage{verbatim}
\usepackage{mathtools}   
\usepackage{mathrsfs} 
\usepackage{xfrac} 
\usepackage{tabu}
\usepackage{color}
\usepackage{bbm}

\definecolor{myred}{RGB}{255, 0, 0}
\definecolor{myblue}{RGB}{0, 0, 255}
\newtheorem{theorem}{Theorem}
\newtheorem{lemma}{Lemma}
\newtheorem{proposition}{Proposition}

\newcommand{\nn}{\nonumber}
\newcommand{\IND}{\mathbbm{1}}
\newcommand {\Exp} {\mathbb{E}}
\newcommand {\prob} {\mathbb{P}}
\newcommand{\DEF}{\overset{\Delta}{=}}

\newcommand{\dfn}{\stackrel{\triangle}{=}}

\newcommand {\lexe} {\stackrel{\cdot} {\leq}}

\newcommand {\bx} {\boldsymbol{x}}
\newcommand {\by} {\boldsymbol{y}}
\newcommand {\bz} {\mbox{\boldmath $z$}}

\newcommand {\bX} {\mbox{\boldmath $X$}}

\newcommand {\bY} {\mbox{\boldmath $Y$}}

\newcommand{\calB}{{\cal B}}
\newcommand{\calC}{{\cal C}}

\newcommand{\calE}{{\cal E}}

\newcommand{\calQ}{{\cal Q}}

\newcommand{\calT}{{\cal T}}

\newcommand{\calX}{{\cal X}}
\newcommand{\calY}{{\cal Y}}

\begin{document}
\thispagestyle{empty}
\title{The MMI Decoder is Asymptotically Optimal for the Typical Random Code and for the Expurgated Code\footnote{
		This research was supported by the Israel Science Foundation (ISF) grant no.\ 137/18.}\\}
\author{\\ Ran Tamir (Averbuch) and Neri Merhav\\}
\maketitle
\begin{center}
	The Andrew \& Erna Viterbi Faculty of Electrical Engineering \\
	Technion - Israel Institute of Technology \\
	Technion City, Haifa 3200003, ISRAEL \\
	\{rans@campus, merhav@ee\}.technion.ac.il
\end{center}
\vspace{1.5\baselineskip}
\setlength{\baselineskip}{1.5\baselineskip}

\begin{abstract}
	We provide two results concerning the optimality of the maximum mutual information (MMI) decoder. First, we prove that the error exponents of the typical random codes under the optimal maximum likelihood (ML) decoder and the MMI decoder are equal. 
	As a corollary to this result, we also show that the error exponents of the expurgated codes under the ML and the MMI decoders are equal.
	These results strengthen the well known result due to Csisz\'ar and K\"orner, according to which, these decoders achieve equal random coding error exponents, since the error exponents of the typical random code and the expurgated code are strictly higher than the random coding error exponents, at least at low coding rates.	
	While the universal optimality of the MMI decoder, in the random-coding error exponent sense, is easily proven by commuting the expectation over the channel noise and the expectation over the ensemble, when it comes to typical and expurgated exponents, this commutation can no longer be carried out. 
	Therefore, the proof of the universal optimality of the MMI decoder must be completely different and it turns out to be highly non-trivial. \\
	
\end{abstract}

\clearpage
\section{Introduction}

The error exponent of the typical random code (TRC) \cite{MERHAV_TYPICAL} is defined as\footnote{Note that this definition is different from the ordinary random coding exponent, which is given by $E_{\mbox{\tiny r}}(R) = \lim_{n \to \infty} \left\{ - \tfrac{1}{n} \log \mathbb{E} \left[P_{\mbox{\tiny e}}(\calC_{n}) \right] \right\}$, where the notations are similar to those in \eqref{TRC_DEF} above.}  
\begin{align} \label{TRC_DEF}
E_{\mbox{\tiny trc}}(R) = \lim_{n \to \infty} \left\{- \tfrac{1}{n} \mathbb{E} \left[\log P_{\mbox{\tiny e}}(\calC_{n}) \right] \right\},
\end{align}
where $R$ is the coding rate, $P_{\mbox{\tiny e}}(\calC_{n})$ is the error probability of a codebook $\calC_{n}$, and the expectation is with respect to (w.r.t.) the randomness of $\calC_{n}$ across the ensemble of codes.

In \cite{BargForney}, Barg and Forney considered TRCs with independently and identically distributed codewords as well as typical linear codes, for the special case of the binary symmetric channel with maximum likelihood (ML) decoding. 
In \cite{PRAD2014} Nazari {\em et al.} provided bounds on the error exponents of TRCs for both discrete memoryless channels (DMC) and multiple--access channels. 
In a recent article by Merhav \cite{MERHAV_TYPICAL}, an exact single--letter expression has been derived for the error exponent of typical, random, fixed composition codes, over DMCs, and a wide class of (stochastic) decoders, collectively referred to as the generalized likelihood decoder (GLD).
Recently, Merhav has studied error exponents of TRCs for the colored Gaussian channel \cite{MERHAV_GAUSS}, typical random trellis codes \cite{MERHAV_TRELLIS}, and has derived a Lagrange--dual lower bound to the TRC exponent \cite{MERHAV_IID}. 
Lately, Tamir {\em et al.} have studied large deviations around the TRC exponent \cite{TMWG}, and finally, Tamir and Merhav have studied error exponents of typical random Slepian--Wolf codes in \cite{TM}.   

Concerning universal decoding for unknown channels, Goppa \cite{GOPPA} was the first to propose the maximum mutual information (MMI) decoder, which decodes the message as the one whose codeword has the largest empirical mutual information with the channel output sequence. Goppa proved that for DMCs, MMI decoding attains capacity. Csisz\'ar and K\"orner \cite[Theorem 5.2]{CK11} have further showed that the random coding error exponent of the MMI decoder, pertaining to the ensemble of the uniform random coding distribution over a certain type class, is equal to the random coding error exponent of the optimum ML decoder.   

In this work, we prove that the error exponents of the TRC under ML and MMI decoding are exactly the same. 
This result improves upon the universal optimality of the MMI decoder proved in \cite{CK11}, since the error exponent of the TRC is strictly higher than the ordinary random coding error exponent, at least at low coding rates \cite{MERHAV_TYPICAL}. 
The fact that the MMI decoder is optimal also w.r.t.\ the TRC is non-trivial, at least not to the authors of this paper. The proof of optimality of the MMI decoder w.r.t.\ the random coding error exponent relies heavily on the possibility to commute the expectations over the channel noise and the randomness of the ensemble of codes. Here, in case of TRCs, this can no longer be done, because, by definition of the TRC exponent, we first apply the logarithmic function on the error probability and only then average over the randomness of the codebook. 
Therefore, the proof of our new result is much more involved than in ordinary random coding.

Universal decoding w.r.t.\ TRCs has already been considered in \cite{TM}. It was proved in \cite{TM} that for Slepian--Wolf source coding, the error exponent of the TRC under the optimal maximum a-posteriori decoder is equal to the TRC exponent under two different universal decoders: the minimum conditional empirical entropy decoder and its stochastic counterpart. While the universality result of \cite{TM} was obtained for some (semi--deterministic) modification of the classic random binning scheme, here, the MMI decoder is proved to be optimal w.r.t.\ the ordinary (fixed--composition) random coding scheme. In light of this difference, we conjecture that for more sophisticated random coding schemes, like the generalized random Gilbert-Varshamov (RGV) code ensemble \cite{SSG}, their TRC exponent under MMI decoding will be even higher.          

Our second result concerns the optimality of MMI decoding w.r.t. expurgated codes. Error exponents of expurgated codes were first developed for the ML decoder \cite{Gal65}, \cite{CKM77}, a few years later for a more general family of deterministic decoders \cite{CKgraph}, and recently for the GLD \cite{MERHAV2017}. In \cite[Section V]{CKgraph}, the question of finding the channels for which the expurgated exponent can be achieved by the minimum entropy decoder (which is equivalent to the MMI decoder under the fixed--composition code ensemble) was left open. 
Here, under the assumption that only the decoder is unaware of the channel statistics, we conclude that the MMI decoder is asymptotically optimal also for the expurgated code. Thanks to the similarity between the expressions of the TRC exponent \cite{MERHAV_TYPICAL} and the expurgated bound \cite{MERHAV2017}, this result immediately follows. 
Since we demonstrate a communication system which is universal only at the decoder side, we conjecture that upon relying on the RGV code, full universality may be attained around any DMC, i.e., one may obtain universality in both the codebook generation process and the channel decoding, while achieving an error exponent as high as $E_{\mbox{\tiny ex}}(R)$.

The remaining part of the paper is organized as follows. 
In Section 2, we establish notation conventions. 
In Section 3, we formalize the model and review some background. 
In Section 4, we provide and discuss the main results of this work, and in Section 5, we prove them.

\section{Notation Conventions}

Throughout the paper, random variables will be denoted by capital letters, specific values they 
may take will be denoted by the corresponding lower case letters, and their alphabets will be 
denoted by calligraphic letters. Random vectors and their realizations will be denoted, 
respectively, by capital letters and the corresponding lower case letters, both in the bold face 
font. Their alphabets will be superscripted by their dimensions. 
For example, the random vector $\bX = (X_{1}, \dotsc , X_{n})$, ($n$ -- positive integer) 
may take a specific vector value $\bx = (x_{1}, \dotsc , x_{n})$ in $\mathcal{X}^{n}$, 
the $n$-th order Cartesian power of $\mathcal{X}$, which is the alphabet of each component of this vector. Sources and channels will be subscripted by the names of the relevant random 
variables/vectors and their conditionings, whenever applicable, 
following the standard notation conventions, e.g., $Q_{X}$, $Q_{Y|X}$, and so on. 
When there is no room for ambiguity, these subscripts will be omitted. For a generic joint 
distribution $Q_{XY} = \{Q_{XY}(x,y), x \in \mathcal{X}, y \in \mathcal{Y} \}$, which will often 
be abbreviated by $Q$, information measures will be denoted in the conventional manner, but
with a subscript $Q$, that is, $H_{Q}(X)$ 
is the marginal entropy of $X$, $H_{Q}(X|Y)$ is the conditional entropy of $X$ given $Y$, 
$I_{Q}(X;Y) = H_{Q}(X) - H_{Q}(X|Y)$ is the mutual information between $X$ and $Y$, 
and so on. 
Logarithms are taken to the natural base.
The probability of an event $\calE$ will be denoted by 
$\prob \{\calE\}$, and the expectation operator with respect to (w.r.t.) a probability distribution $Q$ will be denoted by $\mathbb{E}_{Q}[\cdot]$, where the subscript will often be omitted. 
For two positive sequences $a_{n}$ and $b_{n}$, the notation $a_{n} \doteq b_{n}$ will stand 
for equality in the exponential scale, that is, $\lim_{n \to \infty} \frac{1}{n} 
\log \frac{a_{n}}{b_{n}} = 0$. 
Similarly, $a_{n} \lexe b_{n}$ means that $\limsup_{n \to \infty} (1/n) \log \left(a_{n}/b_{n}\right) \leq 0$, and so on.
The indicator function of an event $\calE$ 
will be denoted by $\IND\{\calE\}$. 
The notation $[x]_{+}$ will stand for $\max \{0, x\}$. 

The empirical distribution of a sequence $\bx \in \mathcal{X}^{n}$, which will 
be denoted by $\hat{P}_{\bx}$, is the vector of relative frequencies, $\hat{P}_{\bx}(x)$, 
of each symbol $x \in \mathcal{X}$ in $\bx$. 
The type class of $\bx \in \mathcal{X}^{n}$, denoted $\calT(\bx)$, 
is the set of all vectors $\bx'$ with $\hat{P}_{\bx'} = \hat{P}_{\bx}$. 
When we wish to emphasize the dependence of the type class on the empirical 
distribution $\hat{P}$, we will denote it by $\calT(\hat{P})$. Information measures 
associated with empirical distributions will be denoted with `hats' and will be subscripted 
by the sequences from which they are induced. For example, the entropy associated 
with $\hat{P}_{\bx}$, which is the empirical entropy of $\bx$, will be denoted 
by $\hat{H}_{\bx}(X)$. Similar conventions will apply to the joint empirical distribution, 
the joint type class, the conditional empirical distributions and the conditional type classes 
associated with pairs (and multiples) of sequences of length $n$. 
Accordingly, $\hat{P}_{\bx\by}$ would be the joint empirical distribution 
of $(\bx, \by) = \{(x_{i}, y_{i})\}_{i=1}^{n}$, $\calT(Q_{X|Y}|\by)$ will stand for the 
conditional type class induced by a sequence $\by$ and a relevant empirical conditional distribution $Q_{X|Y}$, $\hat{I}_{\bx\by}(X;Y)$ will denote the empirical mutual information induced by $\bx$ and $\by$, and so on. 
Similar conventions will apply to triples of sequences, say, $\{(\bx, \by, \bz)\}$, etc. Likewise, when we wish to emphasize the dependence of empirical information measures upon a given empirical distribution given by $Q$, we denote them using the subscript $Q$, as described above.

\section{Problem Setting and Background} \label{SEC3}

\subsection{Problem Setting} \label{SEC3.1}
Consider a DMC $W=\{W(y|x),~x \in \calX,~y \in \calY\}$, where $\calX$ and $\calY$ are the finite input and output alphabets, respectively. When the channel is fed with a sequence $\bx = (x_{1}, \dotsc , x_{n}) \in \calX^{n}$, it produces $\by = (y_{1}, \dotsc , y_{n}) \in \calY^{n}$ according to
\begin{align}
W(\by|\bx) = \prod_{i=1}^{n} W(y_{i}|x_{i}).
\end{align}
Let $\calC_{n}$ be a codebook, i.e., a collection $\{\bx_{0},\bx_{1}, \dotsc, \bx_{M-1}\}$ of $M=e^{nR}$ codewords, $n$ being the block--length and $R$ the coding rate in nats per channel use. When the transmitter wishes to convey a message $m \in \{0,1, \dotsc, M-1\}$, it feeds the channel with $\bx_{m}$.
We assume that messages are chosen with equal probability.
We consider the ensemble of constant composition codes: for a given distribution $Q_{X}$ over $\calX$, all vectors in $\calC_{n}$ are uniformly and independently drawn from the type class $\calT(Q_{X})$.

We consider here two deterministic decoders:
the optimal (MAP) decoder estimates $\hat{m}$, using the channel output $\by$, according to
\begin{align}
\hat{m}(\by) =  \operatorname*{arg\,max}_{m \in \{0,1,\ldots,M-1\}} W(\by | \bx_{m}),
\end{align}
while the MMI decoder estimates $\hat{m}$ according to
\begin{align}
\hat{m}(\by) =  \operatorname*{arg\,max}_{m \in \{0,1,\ldots,M-1\}} \hat{I}_{\bx_{m}\by}(X;Y). 
\end{align}

Let $\bY \in \calY^{n}$ be the random channel output resulting from the transmission of $\bx_{m}$. For a given code $\calC_{n}$, define the error probability as 
\begin{align}
P_{\mbox{\tiny e}}(\calC_{n}) = \frac{1}{M} \sum_{m=0}^{M-1} \prob \{\hat{m}(\bY) \neq m | m~\mbox{sent} \},
\end{align}
where $\prob\{\cdot\}$ designates the probability
measure associated with the randomness of the channel output given its input.

\subsection{Background}
In pure channel coding, 
Merhav \cite{MERHAV_TYPICAL} has derived a single--letter expression for the error exponent of the typical random fixed composition code,
\begin{align}
E_{\mbox{\tiny trc}}(R,Q_{X})
= \lim_{n \to \infty} \left\{- \tfrac{1}{n} \mathbb{E} \left[\log P_{\mbox{\tiny e}}(\calC_{n}) \right] \right\}.
\end{align}
In order to present the main result of \cite{MERHAV_TYPICAL}, we define first a few quantities. 
Define   
\begin{align} \label{a_DEF}
a(R,Q_{Y}) = \max_{\{Q_{\tilde{X}|Y}:~I_{Q}(\tilde{X};Y) \leq R,~ Q_{\tilde{X}}=Q_{X}\}} g(Q_{\tilde{X}Y}),
\end{align}
where either $g(Q)=\mathbb{E}_{Q}[\log W(Y|X)]$ for ML decoding or $g(Q)=I_{Q}(X;Y)$ for MMI decoding. Also define 
\begin{align} \label{Gamma_DEF}
\Gamma(Q_{XX'},R) &= \min_{\{Q_{Y|XX'}:~ g(Q_{X'Y}) \geq \max\{g(Q_{XY}),a(R,Q_{Y})\} \}} \{-\mathbb{E}_{Q}[\log W(Y|X)] - H_{Q}(Y|X,X')\}.
\end{align}
Under the above defined quantities, the error exponent of the TRC is given by \cite{MERHAV_TYPICAL} 
\begin{align} \label{TRCexponent}
E_{\mbox{\tiny trc}}(R,Q_{X})
= \min_{\{Q_{X'|X}:~I_{Q}(X;X') \leq 2R, ~Q_{X'}=Q_{X}\}} \{\Gamma(Q_{XX'},R) + I_{Q}(X;X') - R\}. 
\end{align}

\section{Main Results}

\subsection{Typical Random Codes}
Our main result is the following, which is proved in Section \ref{SEC_PROOF_THM1}. 

\begin{theorem} \label{MMI_TRC}
	For any DMC, the MMI decoder is optimal with respect to the TRC. 
\end{theorem}

As mentioned before, Csisz\'ar and K\"orner \cite[Theorem 5.2]{CK11} have proved that the random coding error exponent of the MMI decoder, pertaining to the ensemble of fixed--composition codes, is as high as the random coding error exponent of the optimum ML decoder. 
The fact that the MMI decoder is also optimal w.r.t.\ the TRC is non-trivial. The proof of optimality of the MMI decoder w.r.t.\ the random coding error exponent relies heavily on the possibility to average directly the error probability, which is defined as
\begin{align}
P_{\mbox{\tiny e}}(\calC_{n}) = \frac{1}{M} \sum_{m=0}^{M-1} \sum_{\by \in \calY^{n}} W(\by|\bx_{m}) \IND \{\hat{m}(\by) \neq m \},
\end{align} 
by first calculating the expectation over the randomness of the ensemble of codes and only then, calculating the expectation over the channel noise. 
Here, when it comes to TRCs, this can no longer be done, because we first apply the logarithmic function on the probability of error and only then average over the randomness of the codebook, and therefore, the proof of Theorem \ref{MMI_TRC} is much more involved than in ordinary random coding.   


Concerning stochastic decoders \cite{MERHAV2017}, let us recall the result of \cite{LCV2017}, which asserts that the probability of error for ordinary likelihood decoding (\cite[Eq.\ (3)]{MERHAV2017}) is at most twice the error probability of ML decoding, which guarantees that the error exponents of the TRC under the ML and the ordinary likelihood decoders are equal. When it comes to universal decoding, a stochastic decoder which is based on the mutual information is strictly suboptimal, as follows by numerical results. As far as we can tell, only deterministic, universal MMI decoding competes well with ML decoding, but not its stochastic counterparts.

\subsection{Expurgated Codes}
The main result of \cite[Section 5]{MERHAV2017} was stated and proved for the GLD. 
The GLD chooses the estimated message $\hat{m}$ according to the following posterior probability mass function, induced by the channel output $\by$:
\begin{align}
\label{StrongGLD}
\prob \left\{ \hat{M}=m \middle| \by \right\}  =
\frac{\exp \{n g( \hat{P}_{\bx_{m} \by } ) \}} {\sum_{m'=0}^{M-1}  
	\exp \{n g( \hat{P}_{\bx_{m'} \by } ) \} } ,
\end{align}
where $\hat{P}_{\bx_{m} \by }$ is the empirical distribution of $(\bx_{m}, \by)$, and $g(\cdot)$ is a given continuous, real--valued functional of this empirical distribution.
The GLD provides a unified framework which covers several important special cases, e.g., matched likelihood decoding, mismatched decoding, ML decoding, and universal decoding (similarly to the $\alpha$--decoders described in \cite{CKgraph}). 
In particular, we recover the ML decoder by choosing the decoding metric
\begin{align}
g(Q_{XY}) = \beta \sum_{x \in \calX} \sum_{y \in \calY} Q_{XY}(x,y) \log W(y|x),
\end{align}
and letting $\beta \to \infty$.
A more detailed discussion is given in \cite{MERHAV2017}.

The proof in \cite{MERHAV2017} was corrected a short time after, concluding that the general expression in \cite{MERHAV2017} is still correct, at least when $g(Q_{XY})$ is an affine functional of $Q_{XY}$, which is the case of the ordinary matched/mismatched stochastic likelihood decoder. Since we need the expurgated exponent to hold for nonlinear decoding metrics as well (e.g., for MMI decoding), we first prove that \cite[Theorem 2]{MERHAV2017} holds for every continuous, real--valued functional $g(Q_{XY})$.    

For a given code $\calC_{n}$, the probability of error given that message $m$ was transmitted is given by
\begin{align}
\label{PROBABILITYofErrorDEF}
P_{\mbox{\tiny e}|m}(\calC_{n})
= \sum_{m' \neq m} \sum_{\by \in \calY^{n}} W(\by|\bx_{m}) \cdot \frac{\exp\{n g(\hat{P}_{\bx_{m'}\by}) \}}{\sum_{\tilde{m}=0}^{M-1} \exp\{n g(\hat{P}_{\bx_{\tilde{m}}\by}) \}}.
\end{align}
In order to characterize the expurgated exponent, we define first a few quantities. Let  
\begin{align}
\alpha(R,Q_{Y}) = \max_{\{Q_{\tilde{X}|Y}:~I_{Q}(\tilde{X};Y) \leq R,~ Q_{\tilde{X}}=Q_{X}\}} \{ g(Q_{\tilde{X}Y}) - I_{Q}(\tilde{X};Y) + R \},
\end{align}
and 
\begin{align}
\tilde{\Gamma}(Q_{XX'},R) &= \min_{Q_{Y|XX'}} \{-\mathbb{E}_{Q}[\log W(Y|X)] - H_{Q}(Y|X,X') \nn \\
&~~~+ [\max\{g(Q_{XY}),\alpha(R,Q_{Y})\} - g(Q_{X'Y})]_{+}\}.
\end{align}
Then, the following proposition is proved in Appendix E.
\begin{proposition} \label{Prop1}
	There exists a sequence of constant composition codes, $\{\calC_{n},~ n=1,2,\ldots\}$, with composition $Q_{X}$, such that
	\begin{align}
	\liminf_{n \to \infty} \left[- \frac{\log \max_{m} P_{\mbox{\tiny e}|m}(\calC_{n})}{n}\right] \geq E_{\mbox{\tiny ex}}(R,Q_{X}),
	\end{align}
	where,
	\begin{align} \label{Expurgated_Exponent}
	E_{\mbox{\tiny ex}}(R,Q_{X})
	= \min_{\{Q_{X'|X}:~I_{Q}(X;X') \leq R, ~Q_{X'}=Q_{X}\}} \{\tilde{\Gamma}(Q_{XX'},R) + I_{Q}(X;X') - R\}.
	\end{align}
\end{proposition}
For ML or MMI decoding, we consider $a(R,Q_{Y})$ and $\Gamma(Q_{XX'},R)$, as defined in \eqref{a_DEF} and \eqref{Gamma_DEF}, respectively, instead of $\alpha(R,Q_{Y})$ and $\tilde{\Gamma}(Q_{XX'},R)$. 

Before stating our main result here, one comment is now in order. One must note that the expurgation process of the codebook relies on the knowledge of the channel statistics, as is evident from the proof in Appendix E. Hence, we assume that only the decoder is ignorant of the channel statistics, while the decoder (or some third party that expurgates the codebook) knows them perfectly. Yet, this assumption can be relaxed by considering more sophisticated code ensembles, like the generalized random Gilbert-Varshamov (RGV) codes \cite{SSG}. The RGV code ensemble is, in fact, inherently expurgated, and it is proved in \cite{SSG} that its random coding error exponent is at least as high as the expurgated exponent derived by Csisz\'ar and K\"orner \cite{CKgraph}. We argue that by relying on the RGV code, one may attain universality (with respect to the channel statistics) in both the codebook generation process and the channel decoding, while achieving an error exponent as given in \eqref{Expurgated_Exponent}. We will not elaborate more on this issue.

Then, our main result is the following. 

\begin{theorem} \label{MMI_EX}
	For any DMC, the MMI decoder is optimal with respect to the expurgated code. 
\end{theorem}

Since the expressions of the TRC exponent \eqref{TRCexponent} and the expurgated bound \eqref{Expurgated_Exponent} are very similar to each other, and differ only in the constraint of the outer minimization, the proof of this theorem is almost identical to the proof of Theorem \ref{MMI_TRC}, and hence omitted.

\section{Proof of Theorem \ref{MMI_TRC}} \label{SEC_PROOF_THM1}
Before proving Theorem \ref{MMI_TRC}, we start with the following series of partial results, that are going to be instrumental in proving Theorem \ref{MMI_TRC}.
In order to present them, we make a few definitions. Let
\begin{align} \label{DEF_G}
G(y,\sigma,\tau,V) = \left(\sum_{x} W(y|x)^{1/\sigma} Q_{X}(x)^{\tau/\sigma} V(x)^{-\tau/\sigma} \right)^{\sigma},
\end{align}
as well as 
\begin{align}
\Psi(Q_{XX'}) = \sup_{s \geq 0} \left\{ - \sum_{x,x'} Q_{XX'}(x,x') \log \left[  \sum_{y} W(y|x)^{1-s} W(y|x')^{s} \right] \right\},
\end{align}
and
\begin{align}
\Theta(Q_{XX'},R) &= \sup_{\rho \geq 0} \inf_{\sigma \geq 0} \inf_{\tau \geq 0} \min_{V} \left\{ \rho \sigma (R-H_{Q}(X)) \right.\nn \\
&\left.~~~- \sum_{x,x'} Q_{XX'}(x,x')  
\log \left[ \sum_{y} W(y|x) W(y|x')^{\rho} G(y,\sigma,\tau,V)^{-\rho}  \right] \right\}.
\end{align}
Also denote 
\begin{align}
\Lambda(Q_{XX'}) = \sup_{\mu \geq 0} \min_{Q_{Y|XX'}}  \left\{-\mathbb{E}_{Q}[\log W(Y|X)] - H_{Q}(Y|X,X') + \mu \left(I_{Q}(X;Y) - I_{Q}(X';Y) \right)\right\},
\end{align}
and
\begin{align} 
\Phi(Q_{XX'},R) = \sup_{\mu \geq 0} \min_{Q_{Y|XX'}}  \left\{-\mathbb{E}_{Q}[\log W(Y|X)] - H_{Q}(Y|X,X') + \mu \left(R - I_{Q}(X';Y) \right)\right\}.
\end{align}
Then, the following lemma is proved in Appendixes A and B:
\begin{lemma} \label{Lemma1}
	The TRC error exponent under ML decoding is upper-bounded by
	\begin{align} \label{Lemma1RES1}
	E_{\mbox{\tiny trc}}^{\mbox{\tiny ML}}(R, Q_{X})
	\leq \min_{\{Q_{X'|X}:~I_{Q}(X;X')\leq 2R, ~Q_{X'}=Q_{X}\}} \{\max\{\Psi(Q_{XX'}),\Theta(Q_{XX'},R)\} + I_{Q}(X;X') - R\}.
	\end{align}
	Furthermore, the TRC error exponent under MMI decoding is lower-bounded by
	\begin{align} \label{Lemma1RES2}
	E_{\mbox{\tiny trc}}^{\mbox{\tiny MMI}}(R, Q_{X})
	\geq \min_{\{Q_{X'|X}:~I_{Q}(X;X')\leq 2R, ~Q_{X'}=Q_{X}\}} \{\max\{\Lambda(Q_{XX'}),\Phi(Q_{XX'},R)\} + I_{Q}(X;X') - R\}.
	\end{align}  
\end{lemma}
The following results are proved in Appendixes C and D: 
\begin{lemma} \label{Lemma3}
	  It holds that 
	  \begin{align} \label{Lemma3RES1}
	  \Psi(Q_{XX'}) \leq \Lambda(Q_{XX'})
	  \end{align}
	  and,
	  \begin{align} \label{Lemma3RES2}
	  \Theta(Q_{XX'},R) \leq \Phi(Q_{XX'},R).
	  \end{align}
\end{lemma}
Finally, we are in a position to compare between $E_{\mbox{\tiny trc}}^{\mbox{\tiny ML}}(R, Q_{X})$ and $E_{\mbox{\tiny trc}}^{\mbox{\tiny MMI}}(R, Q_{X})$:
\begin{align}
	 &E_{\mbox{\tiny trc}}^{\mbox{\tiny ML}}(R, Q_{X}) \nn \\
	 &\leq \min_{\{Q_{X'|X}:~I_{Q}(X;X')\leq 2R, ~Q_{X'}=Q_{X}\}} \{\max\{\Psi(Q_{XX'}),\Theta(Q_{XX'},R)\} + I_{Q}(X;X') - R\} \\
	 &\leq \min_{\{Q_{X'|X}:~I_{Q}(X;X')\leq 2R, ~Q_{X'}=Q_{X}\}} \{\max\{\Lambda(Q_{XX'}),\Phi(Q_{XX'},R)\} + I_{Q}(X;X') - R\} \\
	 &\leq E_{\mbox{\tiny trc}}^{\mbox{\tiny MMI}}(R, Q_{X}),
\end{align}
hence the optimality of MMI decoding follows and Theorem \ref{MMI_TRC} is proved.

\section*{Appendix A}
\renewcommand{\theequation}{A.\arabic{equation}}
\setcounter{equation}{0}  
\subsection*{Proof of eq.\ \eqref{Lemma1RES1} of Lemma \ref{Lemma1}}

First of all, note that 
\begin{align}
&\Gamma(Q_{XX'},R) \nn \\ 
\label{ToCall1}
&= \min_{\{Q_{Y|XX'}:~ g(Q_{X'Y}) \geq \max\{g(Q_{XY}),a(R,Q_{Y})\} \}} \{-\mathbb{E}_{Q}[\log W(Y|X)] - H_{Q}(Y|X,X')\}  \nonumber \\
&= \min_{Q_{Y|XX'}} \sup_{\rho \geq 0} \left\{-\mathbb{E}_{Q}[\log W(Y|X)] - H_{Q}(Y|X,X') + \rho \left(\max\{g(Q_{XY}),a(R,Q_{Y})\} - g(Q_{X'Y}) \right)\right\} \\
&= \sup_{\rho \geq 0} \min_{Q_{Y|XX'}}  \left\{-\mathbb{E}_{Q}[\log W(Y|X)] - H_{Q}(Y|X,X') + \rho \left(\max\{g(Q_{XY}),a(R,Q_{Y})\} - g(Q_{X'Y}) \right)\right\} \\
&= \max \left\{ \sup_{\rho \geq 0} \min_{Q_{Y|XX'}} \{ -\mathbb{E}_{Q}[\log W(Y|X)] - H_{Q}(Y|X,X') + \rho(g(Q_{XY}) - g(Q_{X'Y})) \}, \right. \nn \\
&~~~~~~~~\left. \sup_{\rho \geq 0} \min_{Q_{Y|XX'}} \{ -\mathbb{E}_{Q}[\log W(Y|X)] - H_{Q}(Y|X,X') + \rho(a(R,Q_{Y}) - g(Q_{X'Y})) \} \right\},
\end{align}
since the objective function in \eqref{ToCall1} is convex in $Q_{Y|XX'}$ under ML decoding.
Denote 
\begin{align}
\Psi(Q_{XX'}) = \sup_{\rho \geq 0} \min_{Q_{Y|XX'}} \{ -\mathbb{E}_{Q}[\log W(Y|X)] - H_{Q}(Y|X,X') + \rho(g(Q_{XY}) - g(Q_{X'Y})) \},
\end{align}
and
\begin{align} \label{DEF_XI}
\Xi(Q_{XX'},R) = \sup_{\rho \geq 0} \min_{Q_{Y|XX'}} \{ -\mathbb{E}_{Q}[\log W(Y|X)] - H_{Q}(Y|X,X') + \rho(a(R,Q_{Y}) - g(Q_{X'Y})) \}.
\end{align} 	
Now,
\begin{align}
&\Psi(Q_{XX'}) \nn \\
\label{ToExp2}
&= \sup_{\rho \geq 0} \min_{Q_{Y|XX'}}  \{ -\mathbb{E}_{Q}[\log W(Y|X)] - H_{Q}(Y|X,X') + \rho (g(Q_{XY}) - g(Q_{X'Y})) \} \\
&=  \sup_{\rho \geq 0} \min_{Q_{Y|XX'}} \{ \Exp_{Q}[\log Q_{Y|XX'}(Y|X,X')] +\Exp_{Q}[\log W(Y|X)^{\rho-1}] - \Exp_{Q}[\log W(Y|X')^{\rho}]
\} \\
&=  \sup_{\rho \geq 0} \min_{Q_{Y|XX'}} \left\{ \Exp_{Q}\left[\log \frac{Q_{Y|XX'}(Y|X,X')}{W(Y|X)^{1-\rho} W(Y|X')^{\rho}}\right] \right\} \\
&=  \sup_{\rho \geq 0} \left\{ - \sum_{x,x'} Q_{XX'}(x,x') \log \left[  \sum_{y} W(y|x)^{1-\rho} W(y|x')^{\rho} \right] \right\}.
\end{align}
Next, consider the following
\begin{align}
a(R,Q_{Y}) 
&= \max_{\{Q_{\tilde{X}|Y}:~I_{Q}(\tilde{X};Y) \leq R,~ Q_{\tilde{X}}=Q_{X}\}} g(Q_{\tilde{X}Y})  \\
&= \max_{\{Q_{\tilde{X}|Y}:~ Q_{\tilde{X}}=Q_{X}\}} \inf_{\sigma \geq 0}\{g(Q_{\tilde{X}Y}) + \sigma 
(R-I_{Q}(\tilde{X};Y)) \} \\
&= \max_{\{Q_{\tilde{X}|Y}:~ Q_{\tilde{X}}=Q_{X}\}} \inf_{\sigma \geq 0}\{g(Q_{\tilde{X}Y}) + \sigma (R-H_{Q}(\tilde{X}) + H_{Q}(\tilde{X}|Y)) \} \\
&= \max_{\{Q_{\tilde{X}|Y}:~ Q_{\tilde{X}}=Q_{X}\}} \inf_{\sigma \geq 0}\{g(Q_{\tilde{X}Y}) + \sigma (R-H_{Q}(X) + H_{Q}(\tilde{X}|Y)) \} \\
&= \inf_{\sigma \geq 0} \max_{\{Q_{\tilde{X}|Y}:~ Q_{\tilde{X}}=Q_{X}\}} \{g(Q_{\tilde{X}Y}) + \sigma (R-H_{Q}(X) + H_{Q}(\tilde{X}|Y)) \} \\
&= \inf_{\sigma \geq 0} \left\{ \sigma (R-H_{Q}(X)) + \max_{\{Q_{\tilde{X}|Y}:~ Q_{\tilde{X}}=Q_{X}\}} \{g(Q_{\tilde{X}Y}) + \sigma H_{Q}(\tilde{X}|Y) \} \right\} \\
\label{ToCall3}
&= \inf_{\sigma \geq 0} \left\{ \sigma (R-H_{Q}(X)) + \max_{Q_{\tilde{X}|Y}} \inf_{\tau \geq 0} \{g(Q_{\tilde{X}Y}) + \sigma H_{Q}(\tilde{X}|Y) - \tau D(Q_{\tilde{X}} \| Q_{X}) \} \right\} .
\end{align}
Now,
\begin{align}
-\tau D(Q_{\tilde{X}} \| Q_{X})
&= \tau H_{Q}(\tilde{X}) + \tau \mathbb{E}_{Q}[\log Q_{X}(\tilde{X})] \\
&= \min_{V} \{\tau H_{Q}(\tilde{X}) + \tau D(Q_{\tilde{X}} \| V) \} + \tau \mathbb{E}_{Q}[\log Q_{X}(\tilde{X})] \\
&= \min_{V} \{-\tau \mathbb{E}_{Q}[\log V(\tilde{X})] \} + \tau \mathbb{E}_{Q}[\log Q_{X}(\tilde{X})] ,
\end{align}
hence,
\begin{align}
&\max_{Q_{\tilde{X}|Y}} \inf_{\tau \geq 0} \{g(Q_{\tilde{X}Y}) + \sigma H_{Q}(\tilde{X}|Y) - \tau D(Q_{\tilde{X}} \| Q_{X}) \} \nn \\
&= \max_{Q_{\tilde{X}|Y}} \inf_{\tau \geq 0} \left\{g(Q_{\tilde{X}Y}) + \sigma H_{Q}(\tilde{X}|Y) + \min_{V} \{-\tau \mathbb{E}_{Q}[\log V(\tilde{X})] \} + \tau \mathbb{E}_{Q}[\log Q_{X}(\tilde{X})] \right\}  \\
&= \max_{Q_{\tilde{X}|Y}} \inf_{\tau \geq 0} \min_{V} \left\{\mathbb{E}_{Q}[\log W(Y|\tilde{X})] + \sigma H_{Q}(\tilde{X}|Y) -\tau \mathbb{E}_{Q}[\log V(\tilde{X})] + \tau \mathbb{E}_{Q}[\log Q_{X}(\tilde{X})] \right\}  \\
&\leq \inf_{\tau \geq 0} \min_{V} \max_{Q_{\tilde{X}|Y}} \left\{\mathbb{E}_{Q}[\log W(Y|\tilde{X})] + \sigma H_{Q}(\tilde{X}|Y) -\tau \mathbb{E}_{Q}[\log V(\tilde{X})] + \tau \mathbb{E}_{Q}[\log Q_{X}(\tilde{X})] \right\}  \\
&= -\sup_{\tau \geq 0} \max_{V} \min_{Q_{\tilde{X}|Y}} \left\{ -\mathbb{E}_{Q}[\log W(Y|\tilde{X})] - \sigma H_{Q}(\tilde{X}|Y) +\tau \mathbb{E}_{Q}[\log V(\tilde{X})] - \tau \mathbb{E}_{Q}[\log Q_{X}(\tilde{X})] \right\}  \\
&= -\sup_{\tau \geq 0} \max_{V} \min_{Q_{\tilde{X}|Y}} \left\{
\sum_{y} Q_{Y}(y) \sum_{x} Q_{\tilde{X}|Y}(x|y) \log \left[\frac{ Q_{\tilde{X}|Y}(x|y)^{\sigma}}{W(y|x) Q_{X}(x)^{\tau} V(x)^{-\tau}}\right] \right\} \\
&= - \sigma \sup_{\tau \geq 0} \max_{V} \min_{Q_{\tilde{X}|Y}} \left\{
\sum_{y} Q_{Y}(y) \sum_{x} Q_{\tilde{X}|Y}(x|y) \log \left[\frac{ Q_{\tilde{X}|Y}(x|y)}{W(y|x)^{1/\sigma} Q_{X}(x)^{\tau/\sigma} V(x)^{-\tau/\sigma}}\right] \right\} \\
&= - \sigma \sup_{\tau \geq 0} \max_{V}  \left\{-
\sum_{y} Q_{Y}(y) \sum_{x} Q_{\tilde{X}|Y}(x|y) \log \left[\sum_{x'} W(y|x')^{1/\sigma} Q_{X}(x')^{\tau/\sigma} V(x')^{-\tau/\sigma} \right] \right\} \\
\label{ToCall2}
&= \inf_{\tau \geq 0} \min_{V} \left\{
\sum_{y} Q_{Y}(y) \log \left(\sum_{x} W(y|x)^{1/\sigma} Q_{X}(x)^{\tau/\sigma} V(x)^{-\tau/\sigma} \right)^{\sigma} \right\}.
\end{align}
Let us denote
\begin{align}
G(y,\sigma,\tau,V) = \left(\sum_{x} W(y|x)^{1/\sigma} Q_{X}(x)^{\tau/\sigma} V(x)^{-\tau/\sigma} \right)^{\sigma},
\end{align}
such that substituting \eqref{ToCall2} back into \eqref{ToCall3} yields
\begin{align}
a(R,Q_{Y}) 
&\leq \inf_{\sigma \geq 0} \inf_{\tau \geq 0} \min_{V} \left\{ \sigma (R-H_{Q}(X)) + 
\sum_{y} Q_{Y}(y) \log G(y,\sigma,\tau,V) \right\} \\
\label{ToCall4}
&= \inf_{\sigma \geq 0} \inf_{\tau \geq 0} \min_{V} \left\{ \sigma (R-H_{Q}(X)) + 
\mathbb{E}_{Q}[\log G(Y,\sigma,\tau,V)] \right\}.
\end{align}
Starting now from \eqref{DEF_XI}, we have that
\begin{align} 
&\Xi(Q_{XX'},R) \nn \\ 
&= \sup_{\rho \geq 0} \min_{Q_{Y|XX'}} \{ -\mathbb{E}_{Q}[\log W(Y|X)] - H_{Q}(Y|X,X') + \rho \left[a(R,Q_{Y}) - g(Q_{X'Y})\right] \} \\
&\leq \sup_{\rho \geq 0} \min_{Q_{Y|XX'}} \left\{ -\mathbb{E}_{Q}[\log W(Y|X)] - H_{Q}(Y|X,X') \right.\nn \\
&\left.~~~+ \rho \left[\inf_{\sigma \geq 0} \inf_{\tau \geq 0} \min_{V} \left\{ \sigma (R-H_{Q}(X)) + 
\mathbb{E}_{Q}[\log G(Y,\sigma,\tau,V)] \right\} - g(Q_{X'Y})\right] \right\} \\
&= \sup_{\rho \geq 0} \min_{Q_{Y|XX'}} \inf_{\sigma \geq 0} \inf_{\tau \geq 0} \min_{V} \left\{ -\mathbb{E}_{Q}[\log W(Y|X)] - H_{Q}(Y|X,X') \right.\nn \\
&\left.~~~+ \rho \left[  \sigma (R-H_{Q}(X)) + 
\mathbb{E}_{Q}[\log G(Y,\sigma,\tau,V)] - \mathbb{E}_{Q}[\log W(Y|X')] \right] \right\}\\
&= \sup_{\rho \geq 0} \inf_{\sigma \geq 0} \inf_{\tau \geq 0} \min_{V} \min_{Q_{Y|XX'}} \left\{ -\mathbb{E}_{Q}[\log W(Y|X)] - H_{Q}(Y|X,X') \right.\nn \\
&\left.~~~+ \rho \sigma (R-H_{Q}(X)) + 
\mathbb{E}_{Q}[\log G(Y,\sigma,\tau,V)^{\rho}] - \mathbb{E}_{Q}[\log W(Y|X')^{\rho}] \right\}\\
&= \sup_{\rho \geq 0} \inf_{\sigma \geq 0} \inf_{\tau \geq 0} \min_{V} \min_{Q_{Y|XX'}} \left\{ \rho \sigma (R-H_{Q}(X)) \right.\nn \\
&\left.~~~+ \sum_{x,x'} Q_{XX'}(x,x') \sum_{y} Q_{Y|XX'}(y|x,x') 
\log \left[\frac{Q_{Y|XX'}(y|x,x')}{W(y|x) W(y|x')^{\rho} G(y,\sigma,\tau,V)^{-\rho} }\right] \right\}\\
&= \sup_{\rho \geq 0} \inf_{\sigma \geq 0} \inf_{\tau \geq 0} \min_{V} \left\{ \rho \sigma (R-H_{Q}(X)) \right.\nn \\
&\left.~~~- \sum_{x,x'} Q_{XX'}(x,x')  
\log \left[ \sum_{y} W(y|x) W(y|x')^{\rho} G(y,\sigma,\tau,V)^{-\rho}  \right] \right\},
\end{align} 
which completes the proof of \eqref{Lemma1RES1}.

\section*{Appendix B}
\renewcommand{\theequation}{B.\arabic{equation}}
\setcounter{equation}{0}  
\subsection*{Proof of eq.\ \eqref{Lemma1RES2} of Lemma \ref{Lemma1}}

Under MMI decoding, the error exponent of the TRC is given by 
\begin{align} 
E_{\mbox{\tiny trc}}^{\mbox{\tiny MMI}}(R,Q_{X})
= \min_{\{Q_{X'|X}:~I_{Q}(X;X') \leq 2R, ~Q_{X'}=Q_{X}\}} \{\Omega(Q_{XX'},R) + I_{Q}(X;X') - R\}, 
\end{align}
where,
\begin{align}
&\Omega(Q_{XX'},R) \nn \\
&= \min_{\{Q_{Y|XX'}:~I_{Q}(X';Y) \geq \max\{I_{Q}(X;Y),R\} \}} \{-\mathbb{E}_{Q}[\log W(Y|X)] - H_{Q}(Y|X,X')\} \\
&= \min_{Q_{Y|XX'}} \sup_{\mu \geq 0} \left\{-\mathbb{E}_{Q}[\log W(Y|X)] - H_{Q}(Y|X,X') + \mu \left(\max\{I_{Q}(X;Y),R\} - I_{Q}(X';Y) \right)\right\} \\
&\geq \sup_{\mu \geq 0} \min_{Q_{Y|XX'}}  \left\{-\mathbb{E}_{Q}[\log W(Y|X)] - H_{Q}(Y|X,X') + \mu \left(\max\{I_{Q}(X;Y),R\} - I_{Q}(X';Y) \right)\right\} \\
&\geq \max \left\{ \sup_{\mu \geq 0} \min_{Q_{Y|XX'}}  \left\{-\mathbb{E}_{Q}[\log W(Y|X)] - H_{Q}(Y|X,X') + \mu \left(I_{Q}(X;Y) - I_{Q}(X';Y) \right)\right\}, \right. \nn \\
&~~~~~~~~~~~\left. \sup_{\mu \geq 0} \min_{Q_{Y|XX'}}  \left\{-\mathbb{E}_{Q}[\log W(Y|X)] - H_{Q}(Y|X,X') + \mu \left(R - I_{Q}(X';Y) \right)\right\} \right\}.
\end{align}
Thus,
\begin{align} \label{ToCall7}
\Omega(Q_{XX'},R) \geq \max \left\{\Lambda(Q_{XX'}), \Phi(Q_{XX'},R)\right\}.
\end{align}

\section*{Appendix C}
\renewcommand{\theequation}{C.\arabic{equation}}
\setcounter{equation}{0}  
\subsection*{Proof of eq.\ \eqref{Lemma3RES1} of Lemma \ref{Lemma3}}

Note that 
\begin{align}  
\Lambda(Q_{XX'})
&= \sup_{\mu \geq 0} \min_{Q_{Y|XX'}} \{-\mathbb{E}_{Q}[\log W(Y|X)] - H_{Q}(Y|X,X') + \mu H_{Q}(Y|X') - \mu H_{Q}(Y|X) \} ,
\end{align}
and since
\begin{align}
\mu H_{Q}(Y|X') - \mu H_{Q}(Y|X)
= \max_{V} \min_{V'} \left\{-\mu \Exp_{Q} [\log V'(Y|X')] + \mu \Exp_{Q} [\log V(Y|X)] \right\},
\end{align}
we arrive at
\begin{align} 
\Lambda(Q_{XX'}) 
&= \sup_{\mu \geq 0} \min_{Q_{Y|XX'}} \left\{-\mathbb{E}_{Q}[\log W(Y|X)] - H_{Q}(Y|X,X') \right. \nn \\
&\left. ~~~~+ \max_{V} \min_{V'} \left\{-\mu \Exp_{Q} [\log V'(Y|X')] + \mu \Exp_{Q} [\log V(Y|X)] \right\} \right\} \\
&= \sup_{\mu \geq 0} \min_{Q_{Y|XX'}} \max_{V} \min_{V'} \left\{-\mathbb{E}_{Q}[\log W(Y|X)] - H_{Q}(Y|X,X') \right. \nn \\
&\left. ~~~~-\mu \Exp_{Q} [\log V'(Y|X')] + \mu \Exp_{Q} [\log V(Y|X)] \right\} \\
&\geq \sup_{\mu \geq 0} \max_{V} \min_{V'} \min_{Q_{Y|XX'}} \left\{-\mathbb{E}_{Q}[\log W(Y|X)] - H_{Q}(Y|X,X') \right. \nn \\
&\left. ~~~~-\mu \Exp_{Q} [\log V'(Y|X')] + \mu \Exp_{Q} [\log V(Y|X)] \right\} .
\end{align}
We write the objective function as follows:  
\begin{align}
&-\mathbb{E}_{Q}[\log W(Y|X)] - H_{Q}(Y|X,X') -\mu \Exp_{Q} [\log V'(Y|X')] + \mu \Exp_{Q} [\log V(Y|X)] \nn \\
&= \Exp_{Q}[\log Q_{Y|XX'}(Y|X,X')] -\Exp_{Q}[\log W(Y|X)] -\Exp_{Q}[\log V'(Y|X')^{\mu}]
+\Exp_{Q}[\log V(Y|X)^{\mu}] \\
&= \Exp_{Q}\left[\log \frac{Q_{Y|XX'}(Y|X,X')}{W(Y|X) V'(Y|X')^{\mu} V(Y|X)^{-\mu}}\right] \\
&= \sum_{x,x',y} Q_{XX'Y}(x,x',y) \log \left[ \frac{Q_{Y|XX'}(y|x,x')}{\frac{W(y|x) V'(y|x')^{\mu} V(y|x)^{-\mu}}{\sum_{y'} W(y'|x) V'(y'|x')^{\mu} V(y'|x)^{-\mu} }} \right] \nn \\
&~~~~ - \sum_{x,x',y} Q_{XX'Y}(x,x',y) \log \left[  \sum_{y'} W(y'|x) V'(y'|x')^{\mu} V(y'|x)^{-\mu}\right]\\
&\dfn \sum_{x,x',y} Q_{XX'Y}(x,x',y) \log \left[ \frac{Q_{Y|XX'}(y|x,x')}{B_{Y|XX'}(y|x,x')} \right] \nn \\
&~~~~ - \sum_{x,x'} Q_{XX'}(x,x') \log \left[  \sum_{y} W(y|x) V'(y|x')^{\mu} V(y|x)^{-\mu}\right]\\
\label{AToCall5}
&= D(Q_{Y|XX'}\|B_{Y|XX'}|Q_{XX'}) - \sum_{x,x'} Q_{XX'}(x,x') \log \left[  \sum_{y} W(y|x) V'(y|x')^{\mu} V(y|x)^{-\mu}\right].
\end{align}
Now, minimizing over $Q_{Y|XX'}$ cancels out the first summand in \eqref{AToCall5} and we conclude that:  
\begin{align}
\Lambda(Q_{XX'})
&\geq \sup_{\mu \geq 0} \max_{V} \min_{V'} \left\{ - \sum_{x,x'} Q_{XX'}(x,x') \log \left[  \sum_{y} W(y|x) V'(y|x')^{\mu} V(y|x)^{-\mu}\right] \right\} \\
&= - \inf_{\mu \geq 0} \min_{V} \max_{V'} \left\{\sum_{x,x'} Q_{XX'}(x,x') \log \left[ \sum_{y} W(y|x) V'(y|x')^{\mu} V(y|x)^{-\mu}\right] \right\} .
\end{align}
Note that
\begin{align}
\label{Start_point_0}
&\sum_{y} W(y|x) V'(y|x')^{\mu} V(y|x)^{-\mu} \\
&=\sum_{y} W(y|x) W(y|x')^{\mu} W(y|x)^{-\mu} \left(\frac{V'(y|x')}{W(y|x')}\right)^{\mu} 
\left(\frac{V(y|x)}{W(y|x)}\right)^{-\mu} \\
&=\sum_{y} W(y|x)^{1-\mu} W(y|x')^{\mu} \left(\frac{V'(y|x')}{W(y|x')}\right)^{\mu} 
\left(\frac{V(y|x)}{W(y|x)}\right)^{-\mu} \\
\label{AToExp1}
&\leq \left[\sum_{y} \left(W(y|x)^{1-\mu} W(y|x')^{\mu}\right)^{r}\right]^{1/r} 
\left[\sum_{y} \left(\frac{V'(y|x')}{W(y|x')}\right)^{s\mu}\right]^{1/s} 
\left[\sum_{y} \left(\frac{V(y|x)}{W(y|x)}\right)^{-t\mu}\right]^{1/t} \\
\label{ToCall5}
&\leq \left[\sum_{y} W(y|x)^{1-\mu} W(y|x')^{\mu} \right] 
\left[\sum_{y} \left(\frac{V'(y|x')}{W(y|x')}\right)^{s\mu}\right]^{1/s} 
\left[\sum_{y} \left(\frac{V(y|x)}{W(y|x)}\right)^{-t\mu}\right]^{1/t},
\end{align}
where \eqref{AToExp1} is due to the generalized H\"older inequality with $1/r+1/s+1/t=1$.
Thus,
\begin{align}
&\Lambda(Q_{XX'}) \nn \\
&\geq  - \inf_{\mu \geq 0} \min_{V} \max_{V'} \left\{\sum_{x,x'} Q_{XX'}(x,x') \right. \nn \\
&\left.~~ \times \log \left(\left[ \sum_{y} W(y|x)^{1-\mu} W(y|x')^{\mu} \right] 
\left[\sum_{y} \left(\frac{V'(y|x')}{W(y|x')}\right)^{s\mu}\right]^{1/s} 
\left[\sum_{y} \left(\frac{V(y|x)}{W(y|x)}\right)^{-t\mu}\right]^{1/t}\right) \right\} \\
&= - \inf_{\mu \geq 0} \min_{V} \max_{V'} \left\{ \sum_{x,x'} Q_{XX'}(x,x') \log \left[ \sum_{y} W(y|x)^{1-\mu} W(y|x')^{\mu} \right] \right. \nn \\
&\left. ~~~~~~~~~~~+ \sum_{x,x'} Q_{XX'}(x,x') \log \left[ 
\sum_{y} \left(\frac{V'(y|x')}{W(y|x')}\right)^{s\mu} \right]^{1/s} \right. \nn \\
&\left. ~~~~~~~~~~~+ \sum_{x,x'} Q_{XX'}(x,x') \log \left[  
\sum_{y} \left(\frac{V(y|x)}{W(y|x)}\right)^{-t\mu}\right]^{1/t} \right\}  \\
\label{AToCall6}
&= - \inf_{\mu \geq 0} \left\{\sum_{x,x'} Q_{XX'}(x,x') \log \left[ \sum_{y} W(y|x)^{1-\mu} W(y|x')^{\mu} \right] \right. \nn \\
&\left. ~~~~~~~~~~~+ \max_{V'} \left\{\sum_{x'} Q_{X'}(x') \log \left[ 
\sum_{y} \left(\frac{V'(y|x')}{W(y|x')}\right)^{s\mu} \right]^{1/s} \right\} \right. \nn \\
&\left. ~~~~~~~~~~~+ \min_{V} \left\{\sum_{x} Q_{X}(x) \log \left[ \sum_{y} \left(\frac{V(y|x)}{W(y|x)}\right)^{-t\mu}\right]^{1/t} \right\} \right\} .
\end{align}
We facilitate the expression in \eqref{AToCall6} by choosing $V=W$ instead of minimizing over it. This yields  
\begin{align}
\Lambda(Q_{XX'}) 
&\geq - \inf_{\mu \geq 0} \left\{\sum_{x,x'} Q_{XX'}(x,x') \log \left[ \sum_{y} W(y|x)^{1-\mu} W(y|x')^{\mu} \right] \right. \nn \\
&\left. ~~~~~~~~~~~+ \max_{V'} \left\{\sum_{x'} Q_{X'}(x') \log \left[ 
\sum_{y} \left(\frac{V'(y|x')}{W(y|x')}\right)^{s\mu} \right]^{1/s} \right\} \right. \nn \\
&\left. ~~~~~~~~~~~+ \left\{\sum_{x} Q_{X}(x) \log \left[ \sum_{y} \left(\frac{W(y|x)}{W(y|x)}\right)^{-t\mu} \right]^{1/t} \right\} \right\} \\
&= - \inf_{\mu \geq 0} \left\{\sum_{x,x'} Q_{XX'}(x,x') \log \left[ \sum_{y} W(y|x)^{1-\mu} W(y|x')^{\mu} \right] \right. \nn \\
&\left. ~~~~~~~~~~~+ \max_{V'} \left\{\sum_{x'} Q_{X'}(x') \log \left[ 
\sum_{y} \left(\frac{V'(y|x')}{W(y|x')}\right)^{s\mu} \right]^{1/s} \right\} \right. \nn \\
&\left. ~~~~~~~~~~~+ \left(\sum_{x} Q_{X}(x) \log |\calY|^{1/t} \right) \right\} \\
\label{AToCall8}
&= - \inf_{\mu \geq 0} \left\{\sum_{x,x'} Q_{XX'}(x,x') \log \left[ \sum_{y} W(y|x)^{1-\mu} W(y|x')^{\mu} \right] \right. \nn \\
&\left. ~~~~~~~~~~~+ \max_{V'} \left\{\frac{1}{s}\sum_{x'} Q_{X'}(x') \log \left[ 
\sum_{y} \left(\frac{V'(y|x')}{W(y|x')}\right)^{s\mu} \right] \right\} +  \log |\calY|^{1/t} \right\} .
\end{align}
As for the maximization over the auxiliary channel $V'$, we have the following
\begin{align}
&\max_{V'} \left\{\frac{1}{s}\sum_{x'} Q_{X'}(x') \log \left[ 
\sum_{y} \left(\frac{V'(y|x')}{W(y|x')}\right)^{s\mu} \right] \right\} \nn \\
&\leq \frac{1}{s}\sum_{x'} Q_{X'}(x') \max_{V'(\cdot|x')} \left\{ \log \left[ 
\sum_{y} \left(\frac{V'(y|x')}{W(y|x')}\right)^{s\mu} \right] \right\} \\
\label{AToCall7}
&= \frac{1}{s}\sum_{x'} Q_{X'}(x') \log \left[ \max_{V'(\cdot|x')} \left\{  
\sum_{y} \left(\frac{V'(y|x')}{W(y|x')}\right)^{s\mu}  \right\} \right].
\end{align}
We define the Lagrangian function
\begin{align}
F = \sum_{y} \left(\frac{V'(y|x')}{W(y|x')}\right)^{s\mu} - \lambda \left(\sum_{y} V'(y|x') -1 \right).
\end{align}
Now, differentiating with respect to $V'(y|x')$ yields
\begin{align}
\frac{\partial F}{\partial V'(y|x')} 
=  \frac{(s\mu)V'(y|x')^{s\mu-1}}{W(y|x')^{s\mu}} - \lambda.
\end{align}
The requirement $\partial F/\partial V'(y|x')=0$ is equivalent to 
\begin{align}
V'(y|x')^{s\mu-1} = \lambda (1/s\mu) W(y|x')^{s\mu},
\end{align}
or 
\begin{align}
V'(y|x') = \lambda' W(y|x')^{s\mu/(s\mu-1)},
\end{align}
and thus
\begin{align}
V'(y|x') = \frac{W(y|x')^{s\mu/(s\mu-1)}}{\sum_{y'} W(y'|x')^{s\mu/(s\mu-1)}}.
\end{align}
Substituting it back yields
\begin{align}
\sum_{y} \left(\frac{V'(y|x')}{W(y|x')}\right)^{s\mu}
&= \sum_{y} \left(\frac{W(y|x')^{s\mu/(s\mu-1)}}{W(y|x') \sum_{y'} W(y'|x')^{s\mu/(s\mu-1)}}\right)^{s\mu} \\
&= \sum_{y} \left(\frac{W(y|x')^{1/(s\mu-1)}}{\sum_{y'} W(y'|x')^{s\mu/(s\mu-1)}}\right)^{s\mu} \\
&= \frac{\sum_{y} W(y|x')^{s\mu/(s\mu-1)}}{\left(\sum_{y'} W(y'|x')^{s\mu/(s\mu-1)}\right)^{s\mu}} \\
&= \left(\sum_{y} W(y|x')^{s\mu/(s\mu-1)}\right)^{1-s\mu} .
\end{align}
We continue from \eqref{AToCall7} and get that 
\begin{align}
&\frac{1}{s}\sum_{x'} Q_{X'}(x') \log \left[ \max_{V'(\cdot|x')} \left\{  
\sum_{y} \left(\frac{V'(y|x')}{W(y|x')}\right)^{s\mu}  \right\} \right] \\
&= \frac{1}{s}\sum_{x'} Q_{X'}(x') \log \left[ \left(\sum_{y} W(y|x')^{s\mu/(s\mu-1)}\right)^{1-s\mu} \right] \\
\label{AToCall9}
&= \frac{1-s\mu}{s}\sum_{x'} Q_{X'}(x') \log \left[\sum_{y} W(y|x')^{s\mu/(s\mu-1)} \right]. 
\end{align}
Lower-bounding \eqref{AToCall8} using \eqref{AToCall9} yields 
\begin{align}
\Lambda(Q_{XX'}) 
&\geq - \inf_{\mu \geq 0} \left\{\sum_{x,x'} Q_{XX'}(x,x') \log \left( \sum_{y} W(y|x)^{1-\mu} W(y|x')^{\mu} \right) \right. \nn \\
&\left. ~~~~~~~~~~~+ \frac{1-s\mu}{s}\sum_{x'} Q_{X'}(x') \log \left[\sum_{y} W(y|x')^{s\mu/(s\mu-1)} \right] +  \log |\calY|^{1/t} \right\} .
\end{align}
Optimizing over $s$ and $t$ yields
\begin{align}
\label{AToCall10}
\Lambda(Q_{XX'}) 
&\geq - \inf_{\left\{\substack{r,s,t \in (1,\infty),\\ 1/r+1/s+1/t=1}\right\}} \inf_{\mu \geq 0} \left\{\sum_{x,x'} Q_{XX'}(x,x') \log \left( \sum_{y} W(y|x)^{1-\mu} W(y|x')^{\mu} \right) \right. \nn \\
&\left. ~~~~~~~~~~~+ \frac{1-s\mu}{s}\sum_{x'} Q_{X'}(x') \log \left[\sum_{y} W(y|x')^{s\mu/(s\mu-1)} \right] +  \log |\calY|^{1/t} \right\} \\
&= - \inf_{\mu \geq 0} \inf_{\left\{\substack{r,s,t \in (1,\infty),\\ 1/r+1/s+1/t=1}\right\}} \left\{\sum_{x,x'} Q_{XX'}(x,x') \log \left( \sum_{y} W(y|x)^{1-\mu} W(y|x')^{\mu} \right) \right. \nn \\
&\left. ~~~~~~~~~~~+ \frac{1-s\mu}{s}\sum_{x'} Q_{X'}(x') \log \left[\sum_{y} W(y|x')^{s\mu/(s\mu-1)} \right] +  \log |\calY|^{1/t} \right\} \\
&\geq - \inf_{\mu \geq 0} \lim_{\substack{s \to \infty \\ t \to \infty}} \left\{\sum_{x,x'} Q_{XX'}(x,x') \log \left( \sum_{y} W(y|x)^{1-\mu} W(y|x')^{\mu} \right) \right. \nn \\
&\left. ~~~~~~~~~~~+ \frac{1-s\mu}{s}\sum_{x'} Q_{X'}(x') \log \left[\sum_{y} W(y|x')^{s\mu/(s\mu-1)} \right] +  \log |\calY|^{1/t} \right\} \\
&= - \inf_{\mu \geq 0} \left\{\sum_{x,x'} Q_{XX'}(x,x') \log \left( \sum_{y} W(y|x)^{1-\mu} W(y|x')^{\mu} \right) \right. \nn \\
&\left. ~~~~~~~~~~~-\mu \sum_{x'} Q_{X'}(x') \log \left( \sum_{y} W(y|x') \right) \right\} \\
&= - \inf_{\mu \geq 0} \left\{\sum_{x,x'} Q_{XX'}(x,x') \log \left( \sum_{y} W(y|x)^{1-\mu} W(y|x')^{\mu} \right) \right\} \\
\label{End_point_0}
&= \sup_{\mu \geq 0} \left\{ -\sum_{x,x'} Q_{XX'}(x,x') \log \left( \sum_{y} W(y|x)^{1-\mu} W(y|x')^{\mu} \right) \right\} .
\end{align}
Comparing between $\Psi(Q_{XX'})$ and $\Lambda(Q_{XX'})$ yields 
\begin{align}
\Psi(Q_{XX'})
&= \sup_{\mu \geq 0} \left\{ -\sum_{x,x'} Q_{XX'}(x,x') \log \left( \sum_{y} W(y|x)^{1-\mu} W(y|x')^{\mu} \right) \right\}\\
&\leq \Lambda(Q_{XX'}),
\end{align}
which completes the proof of eq.\ \eqref{Lemma3RES1} of Lemma \ref{Lemma3}.

\section*{Appendix D}
\renewcommand{\theequation}{D.\arabic{equation}}
\setcounter{equation}{0}  
\subsection*{Proof of eq.\ \eqref{Lemma3RES2} of Lemma \ref{Lemma3}}

Notice that
\begin{align} 
&\Phi(Q_{XX'},R) \nn \\
&= \sup_{\mu \geq 0} \min_{Q_{Y|XX'}}  \left\{-\mathbb{E}_{Q}[\log W(Y|X)] - H_{Q}(Y|X,X') + \mu \left(R - I_{Q}(X';Y) \right)\right\} \\
&= \sup_{\mu \geq 0} \min_{Q_{Y|XX'}}  \left\{-\mathbb{E}_{Q}[\log W(Y|X)] - H_{Q}(Y|X,X') + \mu R + \mu H_{Q}(Y|X') - \mu H_{Q}(Y) \right\}.
\end{align}
Since
\begin{align}
\mu H_{Q}(Y|X') - \mu H_{Q}(Y)
= \max_{U} \min_{U'} \left\{-\mu \Exp_{Q} [\log U'(Y|X')] + \mu \Exp_{Q} [\log U(Y)] \right\},
\end{align}
we get that 
\begin{align} 
\Phi(Q_{XX'},R) 
&= \sup_{\mu \geq 0} \min_{Q_{Y|XX'}}  \left\{-\mathbb{E}_{Q}[\log W(Y|X)] - H_{Q}(Y|X,X') + \mu R \right. \nn \\
&\left.~~~~+ \max_{U} \min_{U'} \left\{-\mu \Exp_{Q} [\log U'(Y|X')] + \mu \Exp_{Q} [\log U(Y)] \right\} \right\} \\
&\geq \sup_{\mu \geq 0} \max_{U} \min_{U'} \min_{Q_{Y|XX'}}  \left\{-\mathbb{E}_{Q}[\log W(Y|X)] - H_{Q}(Y|X,X') + \mu R \right. \nn \\
&\left.~~~~-\mu \Exp_{Q} [\log U'(Y|X')] + \mu \Exp_{Q} [\log U(Y)]  \right\} \\
&= \sup_{\mu \geq 0} \left\{ \mu R + \max_{U} \min_{U'} \min_{Q_{Y|XX'}}  \left\{-\mathbb{E}_{Q}[\log W(Y|X)] - H_{Q}(Y|X,X')  \right. \right. \nn \\
& \left. \left.~~~~-\mu \Exp_{Q} [\log U'(Y|X')] + \mu \Exp_{Q} [\log U(Y)]  \right\} \right\} .
\end{align}
As for the innermost minimization, we have that
\begin{align}
&\min_{Q_{Y|XX'}}  \left\{-\mathbb{E}_{Q}[\log W(Y|X)] - H_{Q}(Y|X,X') -\mu \Exp_{Q} [\log U'(Y|X')] + \mu \Exp_{Q} [\log U(Y)]  \right\} \nn \\
&=\min_{Q_{Y|XX'}}  \left\{
\sum_{x,x'} Q_{XX'}(x,x') \sum_{y} Q_{Y|XX'}(y|x,x') \log \left[\frac{Q_{Y|XX'}(y|x,x')}{W(y|x)U'(y|x')^{\mu}U(y)^{-\mu}}\right] \right\} \\
&= -\sum_{x,x'} Q_{XX'}(x,x') \log \left[\sum_{y} W(y|x)U'(y|x')^{\mu}U(y)^{-\mu}\right] ,
\end{align}
hence,
\begin{align} 
&\Phi(Q_{XX'},R) \nn \\ 
&\geq \sup_{\mu \geq 0} \left\{ \mu R + \max_{U} \min_{U'} -\sum_{x,x'} Q_{XX'}(x,x') \log \left[\sum_{y} W(y|x)U'(y|x')^{\mu}U(y)^{-\mu}\right] \right\}\\
&= \sup_{\mu \geq 0} \left\{ \mu R - \min_{U} \max_{U'} \sum_{x,x'} Q_{XX'}(x,x') \log \left[\sum_{y} W(y|x)U'(y|x')^{\mu}U(y)^{-\mu}\right] \right\}.
\end{align}
Now, instead of minimizing over $U$, we lower-bound by choosing
\begin{align}
U^{*}(y) = \sum_{x} W(y|x) Q_{X}(x),
\end{align}
which yields
\begin{align} 
\label{ToCall6}
\Phi(Q_{XX'},R)  
&\geq \sup_{\mu \geq 0} \left\{ \mu R -  \max_{U'} \sum_{x,x'} Q_{XX'}(x,x') \log \left[\sum_{y} W(y|x)U'(y|x')^{\mu} U^{*}(y)^{-\mu} \right] \right\}.
\end{align}
The maximization over $U'$ can be solved by following exactly the same lines as we did earlier (see eqs.\ \eqref{Start_point_0}-\eqref{End_point_0}) for the maximization over $V'$. We conclude that
\begin{align}
\Phi(Q_{XX'},R)  
&\geq \sup_{\mu \geq 0} \left\{ \mu R -   \sum_{x,x'} Q_{XX'}(x,x') \log \left[\sum_{y} W(y|x)W(y|x')^{\mu} U^{*}(y)^{-\mu} \right] \right\}.
\end{align}
Now, in order to compare between $\Theta(Q_{XX'},R)$ and $\Phi(Q_{XX'},R)$, first note that 
\begin{align}
\Theta(Q_{XX'},R) 
&= \sup_{\rho \geq 0} \inf_{\sigma \geq 0} \inf_{\tau \geq 0} \min_{V} \left\{ \rho \sigma (R-H_{Q}(X)) \right.\nn \\
&\left.~~~- \sum_{x,x'} Q_{XX'}(x,x')  
\log \left[ \sum_{y} W(y|x) W(y|x')^{\rho} G(y,\sigma,\tau,V)^{-\rho}  \right] \right\} \\
&\leq \sup_{\rho \geq 0} \min_{\sigma \in [0,1]} \inf_{\tau \geq 0} \min_{V} \left\{ \rho \sigma R - \rho \sigma H_{Q}(X) \right.\nn \\
&\left.~~~- \sum_{x,x'} Q_{XX'}(x,x')  
\log \left[ \sum_{y} W(y|x) W(y|x')^{\rho} G(y,\sigma,\tau,V)^{-\rho}  \right] \right\} \\
&\leq \sup_{\rho \geq 0} \min_{\sigma \in [0,1]} \inf_{\tau \geq 0} \min_{V} \left\{ \rho R - \rho \sigma H_{Q}(X) \right.\nn \\
&\left.~~~- \sum_{x,x'} Q_{XX'}(x,x')  
\log \left[ \sum_{y} W(y|x) W(y|x')^{\rho} G(y,\sigma,\tau,V)^{-\rho}  \right] \right\} .
\end{align}
We continue to upper-bound $\Theta(Q_{XX'},R)$ by making the following choice for $V$:
\begin{align}
V^{*}(x) = \frac{W(y|x)^{1/(\tau+\sigma)}Q_{X}(x)^{\tau/(\tau+\sigma)}}{\sum_{x'} W(y|x')^{1/(\tau+\sigma)}Q_{X}(x')^{\tau/(\tau+\sigma)}},
\end{align}
which provides that
\begin{align}
G(y,\sigma,\tau,V^{*}) = \left(\sum_{x} W(y|x)^{1/(\tau+\sigma)}Q_{X}(x)^{\tau/(\tau+\sigma)}\right)^{\sigma+\tau}.
\end{align}
Substituting it back gives
\begin{align}
&\Theta(Q_{XX'},R) \nn \\
&\leq \sup_{\rho \geq 0} \min_{\sigma \in [0,1]} \inf_{\tau \geq 0} \left\{ \rho R - \rho \sigma H_{Q}(X) \right.\nn \\
&\left.~~~- \sum_{x,x'} Q_{XX'}(x,x')  
\log \left[ \sum_{y} W(y|x) W(y|x')^{\rho} \left(\sum_{x} W(y|x)^{1/(\tau+\sigma)}Q_{X}(x)^{\tau/(\tau+\sigma)}\right)^{-\rho(\sigma+\tau)}  \right] \right\} \\
\label{ToCall8}
&\leq \sup_{\rho \geq 0} \min_{\sigma \in [0,1]} \left\{ \rho R - \rho \sigma H_{Q}(X) \right.\nn \\
&\left.~~~- \sum_{x,x'} Q_{XX'}(x,x')  
\log \left[ \sum_{y} W(y|x) W(y|x')^{\rho} \left(\sum_{x} W(y|x)^{1/(1+\sigma)}Q_{X}(x)^{1/(1+\sigma)}\right)^{-\rho(\sigma+1)}  \right] \right\} \\
\label{ToCall9}
&\leq \sup_{\rho \geq 0} \left\{ \rho R - \sum_{x,x'} Q_{XX'}(x,x')  
\log \left[ \sum_{y} W(y|x) W(y|x')^{\rho} \left(\sum_{x} W(y|x)Q_{X}(x)\right)^{-\rho}  \right] \right\} \\
&= \sup_{\rho \geq 0} \left\{ \rho R - \sum_{x,x'} Q_{XX'}(x,x')  
\log \left[ \sum_{y} W(y|x) W(y|x')^{\rho} U^{*}(y)^{-\rho}  \right] \right\} \\
&\leq \Phi(Q_{XX'},R),
\end{align}
where \eqref{ToCall8} follows from the choice $\tau=1$ and \eqref{ToCall9} from the choice $\sigma=0$.

\section*{Appendix E}
\renewcommand{\theequation}{E.\arabic{equation}}
\setcounter{equation}{0}  
\subsection*{Proof of Proposition \ref{Prop1}}

Assuming that message $m$ was transmitted, the probability of error, for a given code $\calC_{n}$, is given by
\begin{align}
P_{\mbox{\tiny e}|m}(\calC_{n})
= \sum_{m' \neq m} \sum_{\by \in \calY^{n}} W(\by|\bx_{m}) \cdot \frac{\exp\{n g(\hat{P}_{\bx_{m'}\by}) \}}{\exp\{n g(\hat{P}_{\bx_{m}\by}) \} + \sum_{\tilde{m} \neq m} \exp\{n g(\hat{P}_{\bx_{\tilde{m}}\by}) \}}.
\end{align}
Let
\begin{align}
\label{Z_DEF}
Z_{m}(\by) = \sum_{\tilde{m} \neq m} \exp\{n g(\hat{P}_{\bx_{\tilde{m}}\by}) \},
\end{align}
fix $\epsilon>0$ arbitrarily small, and for every $\by \in \calY^{n}$, define the set
\begin{align}
\label{B_DEF}
\calB_{\epsilon}(m,\by) = \left\{\calC_{n}:~ Z_{m}(\by) \leq \exp\{n \alpha(R-\epsilon, \hat{P}_{\by})\}  \right\}.
\end{align}
Following the result of \cite[Appendix B]{MERHAV2017}, we know that, considering the ensemble of randomly selected constant composition codes of type $Q_{X}$,
\begin{align}
\label{B_DE_UB}
\prob \{\calB_{\epsilon}(m,\by)\} \leq \exp\{-e^{n\epsilon} + n\epsilon + 1\},
\end{align} 
for every $m \in \{0,1,\dotsc,M-1\}$ and $\by \in \calY^{n}$, and so, by the union bound,
\begin{align}
\label{B_UNION_DEF}
\prob \left\{\bigcup_{\by \in \calY^{n}}\calB_{\epsilon}(m,\by)\right\}
\DEF  \prob \left\{\calB_{\epsilon}(m)\right\} 
&\leq \sum_{\by \in \calY^{n}} \prob \left\{\calB_{\epsilon}(m,\by)\right\} \\
&\leq \sum_{\by \in \calY^{n}} \exp\{-e^{n\epsilon} + n\epsilon + 1\} \\
&= |\calY|^{n} \cdot \exp\{-e^{n\epsilon} + n\epsilon + 1\},
\end{align}
which still decays double--exponentially fast. 
Define the set $\calQ(Q_{X}) = \{Q_{X'|X}:~Q_{X'}=Q_{X}\}$ and the enumerator
\begin{align}
N_{m}(Q_{X'|X}) = \sum_{m' \neq m} \IND \left\{\bx_{m'} \in \calT(Q_{X'|X}|\bx_{m}) \right\}.
\end{align}
Now, for $\rho \geq 1$,
\begin{align}
&\Exp \left[P_{\mbox{\tiny e}|m}(\calC_{n})^{1/\rho} \middle| \bx_{m} \right] \nn \\
&= \Exp \left[P_{\mbox{\tiny e}|m}(\calC_{n})^{1/\rho} \cdot \IND\{\calB_{\epsilon}(m)^{\mbox{\tiny c}} \} \middle| \bx_{m} \right] \nn \\
&~~+ \Exp \left[P_{\mbox{\tiny e}|m}(\calC_{n})^{1/\rho} \cdot \IND\{\calB_{\epsilon}(m) \} \middle| \bx_{m} \right] \\
&= \Exp \left[\left(\sum_{m' \neq m} \sum_{\by \in \calY^{n}} W(\by|\bx_{m}) \cdot \frac{\exp\{n g(\hat{P}_{\bx_{m'}\by}) \}}{\exp\{n g(\hat{P}_{\bx_{m}\by}) \} + Z_{m}(\by)}\right)^{1/\rho} \cdot \IND\{\calB_{\epsilon}(m)^{\mbox{\tiny c}} \} \middle| \bx_{m} \right] \nn \\
&~~+ \prob \left\{\calB_{\epsilon}(m) \middle| \bx_{m} \right\} \\
&\leq \Exp \left[\left(\sum_{m' \neq m} \sum_{\by \in \calY^{n}} W(\by|\bx_{m}) \cdot \min \left\{ 1,\frac{\exp\{n g(\hat{P}_{\bx_{m'}\by}) \}}{\exp\{n g(\hat{P}_{\bx_{m}\by}) \} + \exp\{n \alpha(R-\epsilon, \hat{P}_{\by})\}} \right\} \right)^{1/\rho} \middle| \bx_{m} \right] \nn \\
&~~+ \prob \left\{\calB_{\epsilon}(m) \right\} \\
&\doteq \Exp \left[\left(\sum_{m' \neq m} \exp\{-n \tilde{\Gamma}(\hat{P}_{\bx_{m}\bx_{m'}},R) \} \right)^{1/\rho} \middle| \bx_{m} \right] + \prob \left\{\calB_{\epsilon}(m) \right\} \\
&\leq \Exp \left[\left(\sum_{Q_{X'|X}\in \calQ(Q_{X})} N_{m}(Q_{X'|X}) \cdot \exp\{-n \tilde{\Gamma}(Q_{XX'},R) \} \right)^{1/\rho} \middle| \bx_{m} \right] \nn \\
&~~+ |\calY|^{n} \cdot \exp\{-e^{n\epsilon} + n\epsilon + 1\} \\
\label{ToCall10}
&\lexe \sum_{Q_{X'|X}\in \calQ(Q_{X})} \Exp \left[ N_{m}(Q_{X'|X})^{1/\rho} \middle| \bx_{m} \right] \cdot \exp\{-n \tilde{\Gamma}(Q_{XX'},R)/\rho \} .
\end{align}
The conditional expectation in \eqref{ToCall10} is given by
\begin{align}
\Exp \left[ N_{m}(Q_{X'|X})^{1/\rho} \middle| \bx_{m} \right] &\doteq \left\{   
\begin{array}{l l}
\exp\{n (R - I_{Q}(X;X'))/\rho\}    & \quad I_{Q}(X;X') \leq R   \\
\exp\{n (R - I_{Q}(X;X')) \}    & \quad I_{Q}(X;X') > R   \\
\end{array} \right. \\
&\dfn \exp \{n E(R,Q,\rho)\}. 
\end{align} 
Note that the expression of $E(R,Q,\rho)$ is independent of $\bx_{m}$. Substituting it back into \eqref{ToCall10} provides an upper bound on $\Exp \left[P_{\mbox{\tiny e}|m}(\calC_{n})^{1/\rho} \middle| \bx_{m} \right]$, which is independent of $\bx_{m}$, hence, it also holds for the unconditional expectation, i.e.,
\begin{align}
&\Exp \left[P_{\mbox{\tiny e}|m}(\calC_{n})^{1/\rho} \right] \lexe \sum_{Q_{X'|X}} \exp \{n E(R,Q,\rho)\} \cdot \exp\{-n \tilde{\Gamma}(Q_{XX'},R)/\rho \} \dfn \mathbf{\Delta}.
\end{align}
According to Markov's inequality, we get
\begin{align}
\prob \left\{ \frac{1}{M} \sum_{m=0}^{M-1} P_{\mbox{\tiny e}|m}(\calC_{n})^{1/\rho} > 2 \mathbf{\Delta} \right\} \leq \frac{1}{2},
\end{align}
which means that there exists a code with
\begin{align}
\frac{1}{M} \sum_{m=0}^{M-1} P_{\mbox{\tiny e}|m}(\calC_{n})^{1/\rho} \leq 2 \mathbf{\Delta}.
\end{align}
We conclude that there exists a code $\calC'_{n}$ with $M/2$ codewords for which 
\begin{align}
\max_{m} P_{\mbox{\tiny e}|m}(\calC'_{n})^{1/\rho} \leq 4 \mathbf{\Delta},
\end{align}
and so
\begin{align}
\max_{m} P_{\mbox{\tiny e}|m}(\calC'_{n}) 
&\leq \left(\sum_{Q_{X'|X}\in \calQ(Q_{X})} \exp \{n E(R,Q,\rho)\} \cdot \exp\{-n \tilde{\Gamma}(Q_{XX'},R)/\rho \}\right)^{\rho}\\
&\doteq \sum_{Q_{X'|X}\in \calQ(Q_{X})} \exp \{n \rho E(R,Q,\rho)\} \cdot \exp\{-n \tilde{\Gamma}(Q_{XX'},R) \}\\
&\doteq \exp \left\{-n \cdot \min_{Q_{X'|X}\in \calQ(Q_{X})} [\tilde{\Gamma}(Q_{XX'},R) - \rho E(R,Q,\rho)]\right\},
\end{align}
thus,
\begin{align}
\liminf_{n \to \infty} - \frac{1}{n} \log \max_{m} P_{\mbox{\tiny e}|m}(\calC'_{n}) 
\geq \min_{Q_{X'|X}\in \calQ(Q_{X})} [\tilde{\Gamma}(Q_{XX'},R) - \rho E(R,Q,\rho)].
\end{align}
Since it holds for every $\rho \geq 1$, the negative exponential rate of the maximal probability of error can be bounded as
\begin{align}
\liminf_{n \to \infty} - \frac{1}{n} \log \max_{m} P_{\mbox{\tiny e}|m}(\calC'_{n}) 
&\geq \sup_{\rho \geq 1} \min_{Q_{X'|X}\in \calQ(Q_{X})} [\tilde{\Gamma}(Q_{XX'},R) - \rho E(R,Q,\rho)] \\
&\geq \lim_{\rho \to \infty} \min_{Q_{X'|X}\in \calQ(Q_{X})} [\tilde{\Gamma}(Q_{XX'},R) - \rho E(R,Q,\rho)].
\end{align}
Since
\begin{align}
\lim_{\rho \to \infty} \rho E(R,Q,\rho)
= \left\{   
\begin{array}{l l}
R - I_{Q}(X;X')  & \quad I_{Q}(X;X') \leq R   \\
- \infty         & \quad I_{Q}(X;X') > R   \\
\end{array} \right.  ,
\end{align}
we finally arrive at 
\begin{align}
\liminf_{n \to \infty} - \frac{1}{n} \log \max_{m} P_{\mbox{\tiny e}|m}(\calC'_{n}) 
&\geq \min_{\{Q_{X'|X}:~I_{Q}(X;X') \leq R, ~Q_{X'}=Q_{X}\}} \{\tilde{\Gamma}(Q_{XX'},R) + I_{Q}(X;X') - R\},
\end{align}
and the proof of Proposition \ref{Prop1} is now complete.

\end{document}